\documentclass[prb,
superscriptaddress,showpacs,amsmath,amssymb]{revtex4}
\usepackage{amsfonts}
\usepackage{bm}
\usepackage{verbatim}

\usepackage{graphicx}

\begin{document}

\title{Topological charge of fermions and Landau theory of Fermi liquid}

\date{\today}

\begin{abstract}

In the fermionic liquids, the Fermi surface is topologically stable,\cite{Volovik2003} which is at the origin of the applicability of the Landau theory of Fermi liquid (LFL). The LFL exists under special condition, when the Green's function has a pole with nonzero residue $Z$. Otherwise one has non-Landau Fermi liquid (NLFL), such as Luttinger liquid, which is described by the same topological invariant. It appears that in general this topological invariant is the property of the fermionic particle, i.e. the particle charge (or the electric charge of electron) is equivalent to the topological charge of the fermion. The conservation of the fermionic charge is equivalent to the conservation of the topological charge.
We consider the application of this topological charge  to the Landau theory of Fermi liquids. We also consider the application to non-Fermi liquids and crystalline insulators in relation to the Luttinger theorem.  
\end{abstract}
\pacs{
}

\author{G.E.~Volovik}
\affiliation{Landau Institute for Theoretical Physics, acad. Semyonov av., 1a, 142432,
Chernogolovka, Russia}

\maketitle

\tableofcontents

\section{Introduction} 
 
 Topology in the momentum-frequency $({\bf p},\omega)$ space plays an important role in the classification of topological matter. Here we consider how this topology justifies the Landau theory of Fermi liquids and the Luttinger theorem. We will that each electron state in Fermi liquid, which is characterized by momentum ${\bf p}$, has its own topological invariant $N_\omega ({\bf p})$ with values 0 or 1. This invariant plays the role of the occupation number $n({\bf p})$ of quasiparticles in the Landau theory of Fermi liquid. The total topological invariant of Fermi liquid is equal to the total number of particles in the system. This supports the main conjecture of the Landau theory, that the number of  quasiparticles coincides with the number of particles, and also supports the Luttinger theorem. Due to topological stability, the Luttinger theorem remains valid, even if due to electron-electron interaction the pole in Green's function transforms to zero.
 
 The Fermi surface is also topologically stable, being the border between the regions with different values of the topological invariant $N_\omega ({\bf p})$. We also discuss the Khodel-Shaginyan mechanism of the transformation of the Fermi surface to the flat band due to electron-electron interactions and possibility of the room temperature superconductivity due to extremely large density of states in the flat band. Some experiments, where hints of room-$T$ superconductivity are observed, are mentioned.
 
 The Luttinger theorem is extended to the topological insulators. Also different momentum-space topological invariants, which characterize topological insulators, are discussed together with the corresponding actions of the Chern-Simons type. These actions contain the gauge field $A_{\mu}$ and the translational gauge fields in terms of the elasticity tetrads $E^a_\mu$. This also includes the 
 $\Theta$-term, which represents the so-called strong $CP$ problem in quantum chromodynamics.
 
   \section{Topology of Landau Fermi liquid} 
\label{LandauSec}

   \subsection{Spectral asymmetry index and topological invariant} 

Let us consider first the single particle Hamiltonian ${\cal H}$ and its eigenvalues $E_n$.
The spectral asymmetry index is
\begin{eqnarray}
\nu=-\frac{1}{2}\sum _n {\rm sign} ~ E_n~.
\label{SpectrumAsymmetryIndex0}
\end{eqnarray}
It is the difference between the number of negative and positive energy levels of the Hamiltonian.

This can be written as:
\begin{eqnarray}
-\frac{1}{2}\sum _n {\rm sign} \, E_n=-\sum _n  \int_{-\infty}^{+\infty} \frac{d p_0}{2\pi}
 \frac{E_n}{p_0^2 + E_n^2} =-{\bf Tr} \int_{-\infty}^{+\infty}\frac{d p_0}{2\pi}  \frac{\cal H}{p_0^2 +
{\cal H}^2} \,.
\label{SpectrumAsymmetryIndex00}
\end{eqnarray}
Here $ip_0$ is the imaginary frequency; the integration over $p_0$ is from $-\infty$ to $+\infty$; and ${\bf Tr}$ is over the matrix indices $H_{mn}$.

 The spectral asymmetry index $\nu$ can be expressed in terms of the Green's function,
${\cal G}^{-1}(p_0) = i p_0 - {\cal H}$:
\begin{eqnarray}
\nu ={\bf Tr}  \int_{-\infty}^{+\infty} \frac{d p_0}{2\pi}{\cal G} \,.
\label{SpectrumAsymmetryGreen}
\end{eqnarray}

   \begin{figure}
\centerline{\includegraphics[width=0.5\linewidth]{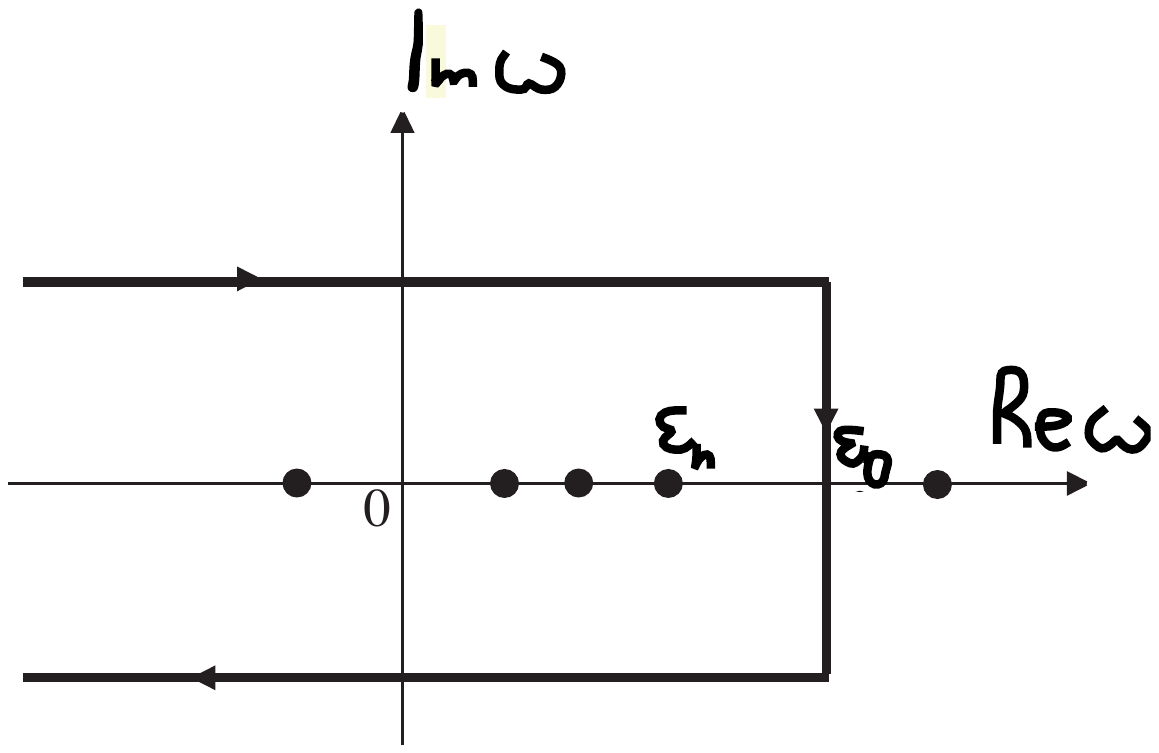}}
\caption{Topological invariant $\nu_\omega$ in Eq.(\ref{Nofmu}) is the integral over the contour  in the $\omega$ plane, $({\rm Re}\, \omega, {\rm Im}\, \omega=p_0)$, shown in the Figure.
The points show the energy levels $\epsilon_n$. The integral over this contour gives the total number of the energy states below the chemical potential (or below the Fermi level $\epsilon_0$ in general). At large negative ${\rm Re}\, \omega$ the contour is effectively closed, and the integer topological charge is well defined.
The integer valued invariant  $\nu_\omega$ experiences jumps -- the topological quantum phase transitions -- when one of the points crosses the Fermi level.}  
\label{SpectrumAss} 
\end{figure}

The next step is to introduce the topological invariant. Let is rewrite Eq.(\ref{SpectrumAsymmetryGreen}) in the following form:
\begin{eqnarray}
\nu={\bf Tr}\, \int  \frac{d p_0}{2\pi i} {\cal G} \partial_{p_0} {\cal G}^{-1} \,.
\label{SpectrumAsymmetryTop}
\end{eqnarray}
When the integral is again along $p_0$ from $-\infty$ to $+\infty$, it is the same as the spectrum asymmetry index in Eq.(\ref{SpectrumAsymmetryGreen}).
$\nu$ becomes the integer valued topological  invariant only if the contour of integration is closed. For that let us choose the contour $C(\omega)$ in Fig. \ref{SpectrumAss}, where the real component of frequency  ${\rm Re}\,\omega$
varies from $-\infty$ to the chemical potential $\mu$ (or $\epsilon_0$) in the considered system:
 \begin{equation}
\nu_\omega={\bf Tr}~\oint_{C(\omega)} {d\omega\over 2\pi i} {\cal G}\partial_\omega {\cal G}^{-1} \,.
\label{Nofmu}
\end{equation}

 For ${\rm Re}\,\omega=-\infty$ there are no energy states, if the energy is restricted from below, and thus the contour  is effectively closed.
That is why $\nu_\omega$ is the total number of states with energies below the chemical potential $\mu$ (or below the Fermi level in general). 
 In case of the fermionic system, where at $T=0$ all the negative energy states are occupied below the Fermi level, then the topological charge $\nu_\omega$ becomes the total number of fermions in the system.
 
 \subsection{Topological invariant and generalization of Luttinger theorem} 

Let us now consider the translational invariant Fermi liquids, where the energy states can be characterized by momentum ${\bf p}$. Then for each state with momentum ${\bf p}$ one introduce the topological invariant  $N_\omega({\bf p})$. 
The invariant 
$N_\omega({\bf p})$ is determined for each momentum ${\bf p}$
\begin{eqnarray}
N_\omega({\bf p})={\bf tr}\, \int_{C}  \frac{d \omega}{2\pi i} \,{\cal G}(\omega,{\bf p}) \partial_{\omega} {\cal G}^{-1}(\omega,{\bf p}) \,,
\label{insulator2}
\end{eqnarray}
Here trace is over all the states with a given momentum ${\bf p}$ (there can be spin degrees of freedom, or bands in case of crystals). For crystalline insulators this invariant was  introduced in Ref.\cite{NissinenHeikkilaVolovik2021}, see Sec.\ref{InsulatorsSec}. 

Eq. (\ref{Nofmu}) can be applicable to the interacting systems too. For a single component interacting Fermi liquids this invariant takes values $0$ and $1$, which corresponds to the empty and occupied states correspondingly. The total topological charge is obtained by summation of the topological charges of fermions:
\begin{equation}
N=\sum_{\bf p} N_\omega({\bf p})\,,
\label{TotalCharge}
\end{equation}
It corresponds to the total number of fermions below the Fermi level, and thus $N$ represents the total number of fermions in the system. In this sense, the number of fermions $N$ in the Fermi liquid is the topological charge.

 Since the integer valued topological invariant is valid for the interacting system, its integer value can be varied only by jump. The interacting system can be obtained from the noninteracting by the adiabatic transformations, until the topological quantum phase transition takes place. Before such transition occurs, the spectral flow is absent, and the value of the topological invariant does not change. In this case Eq. (\ref{Nofmu})  represents the proper generalization of the Luttinger theorem.

 \subsection{Landau phenomenology of Fermi liquids}

   \begin{figure}
\centerline{\includegraphics[width=0.5\linewidth]{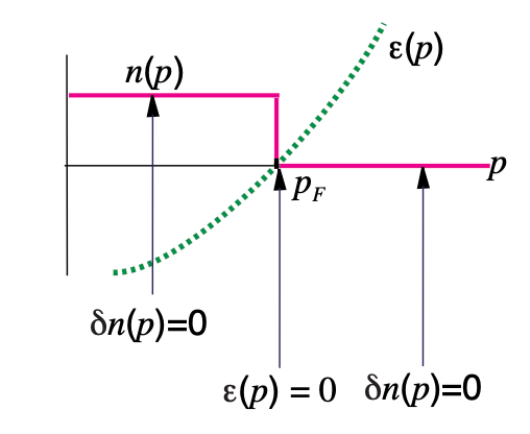}}
\caption{Distribution of fermions in Landau theory. The role of the particle number $n({\bf p})$ is played by topological invariant $N_\omega({\bf p})$ in Eqs. (\ref{insulator2}) and (\ref{LandauOcccupation}), while the real distributions $n({\bf p})$ is illustrated in Fig.\ref {Migdal1}. The topological stability of the Fermi surface, which separates regions with different values of the topological charge $N_\omega({\bf p})$, is supported by the topological invariant $N_1$ in Eq.(\ref{InvariantForFS}).
}  
\label{Migdal2} 
\end{figure}

\begin{figure}
\centerline{\includegraphics[width=0.5\linewidth]{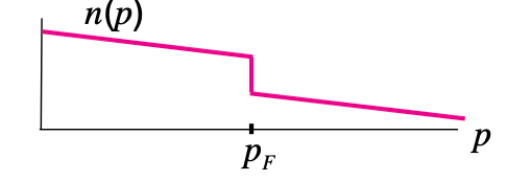}}
\caption{Illustration of the real occupation number $n({\bf p})=<a_{\bf p}^\dagger a_{\bf p}>$  in the Fermi liquid with Migdal jump at the Fermi surface. The Fermi surface is robust to interaction, being supported by the invariant $N_1$, although the singularity at the Fermi surface can be without the Migdal jump.  The invariant $N_1$ is the momentum-space analogue of the circulation number, which characterizes quantized vortices in superfluids. The Luttinger theorem is also robust to interaction. It is valid even if the pole in the Green's function transforms to zero, since this transformation does not change the topological charge $N_\omega({\bf p})$.}  
\label{Migdal1} 
\end{figure}

In case of a single band, one can identify the topological invariant $N_\omega({\bf p})$  with the occupation number in the Landau theory of Fermi liquid, which takes values $0$ and $1$:
  \begin{equation}
N_\omega({\bf p})\equiv n({\bf p}) \,.
\label{LandauOcccupation}
\end{equation} 
This is the famous Landau identification of the number of quasiparticles with the number of particles.
This takes place, because both quasiparticles in interacting systems and the bare electrons are described by the same topological invariant $N_\omega({\bf p})$. This is different from the real distribution of fermions $<a_{\bf p}^\dagger a_{\bf p}>$ in the interacting Fermi liquids with Migdal jump at the Fermi surface, which is illustrated in Fig.(\ref{Migdal1}).

 Example is the topological action which contains the topological invariant $N_\omega({\bf p})$ and gauge field $A_\mu$:
\begin{equation}
 S  = 2\int  d^4x\,  \frac{d^3p}{(2\pi)^3} \, N_\omega({\bf p}) (A_0 + {\bf v}\cdot {\bf A}) \,.
\label{action}
\end{equation} 
Here ${\bf v}$ is the velocity of the Landau electronic liquid and the factor 2 reflects the spin degrees of freedom. The other types of topological actions will be considered later.

Eq.(\ref{action}) gives the charge density (or particle density in electrically neutral Fermi liquid):
\begin{equation}
J^0=\frac{\delta S}{\delta A_0} = 2 \int   \frac{d^3p}{(2\pi)^3} \, N_\omega({\bf p}) \equiv  2 \int   \frac{d^3p}{(2\pi)^3} \, n({\bf p}) \,,
\label{particleDensity}
\end{equation} 
and particle current:
\begin{equation}
{\bf J}= \int  \frac{d^3p}{(2\pi)^3} \, {\bf v} \,N_\omega({\bf p}) =  \int_{N_\omega({\bf p})=1}  \frac{d^3p}{(2\pi)^3}\,{\bf v}  \equiv   \int   \frac{d^3p}{(2\pi)^3} \, n({\bf p}) {\bf v}\,.
\label{LuttingerCurrent}
\end{equation} 

 The identification $N_\omega({\bf p})\equiv n({\bf p})$ at $T=0$ supports the  Landau theory of Fermi liquid, which operates with the function $n({\bf p})$, considered as occupation number for quasiparticles.  
 
\subsection{Topological invariant for Fermi surface} 
  
Let us now consider the border between the states with the topological invariant $N_\omega({\bf p})=1$ and that with  $N_\omega({\bf p})=0$. This is the singular surface in momentum space -- the the Fermi surface. This surface is also described by the momentum space invariant
$N_1$, which is responsible for the topological
stability of the Fermi surface:\cite{Volovik2003}
\begin{equation}
N_1={\bf Tr}~\oint_C {dl\over 2\pi i} {\cal G}(\omega,{\bf p})\partial_l {\cal
G}^{-1}(\omega,{\bf p})~.
\label{InvariantForFS}
\end{equation}
Here the integral is taken over a small contour $C$, which encloses the element of the Fermi
surface, see Fig. 8.1 in Ref. \cite{Volovik2003}.
  
 Equations  (\ref{TotalCharge}) and (\ref{InvariantForFS}) are applicable to the interacting systems as well, which 
represents the proper generalization of the Luttinger theorem\cite{Luttinger1960a,Luttinger1960b}  to the case with several fermionic species.
In this general form the Luttinger theorem can be valid 
not only for metals. Since pole and zero in the Green's function have the same topological invariant $N_1$, this equation can be applicable also for the Mott insulators, which are characterized by zeroes in the Green's function instead of poles.\cite{Dzyaloshinskii2003,Tsvelik2009,Farid2008}

\subsection{From Landau Fermi liquid to flat band} 
\label{FlatBandSec}

   \begin{figure}
\centerline{\includegraphics[width=0.5\linewidth]{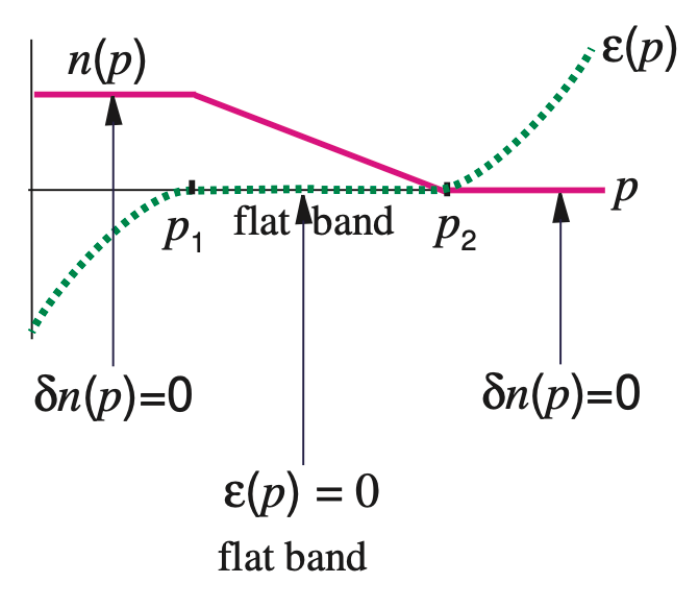}}
\caption{Illustration of Landau theory for Fermi liquid which contains the flat band. Variation of the phenomenological Landau functional gives two solutions: $\delta n({\bf p})=0$ and $\epsilon({\bf p})=0$. The latter is either the Fermi surface in Fig. \ref{Migdal2} or the flat band.}  
\label{FlatB} 
\end{figure}

It was shown by Khodel and Shaginyan that when the interaction between the electrons is sufficiently strong, the Landau theory gives rise to the formation of the flat band in Fig.(\ref{FlatB}).\cite{KhodelShaginyan1990,Volovik1991} The flat band (originally introduced as the Fermi condensate) is the region in the momentum space, where all quasiparticles have zero energy. This phenomenon is also called the “level merging”.\cite{Dolgopolov2022} Formation of the Fermi-ball (the whole ball in momentum space occupied by the zero energy states) has been also obtained using the AdS/CFT correspondence.\cite{FermiBall} 

The Landau theory uses the phenomenological energy functional, whose variation gives the energy spectrum of quasiparticles, $\delta {\cal E}= \sum_{\bf p} \epsilon_{\bf p} \delta n_{\bf p}$.
Since the quasiparticle distribution function is constrained by the Pauli principle $0\leq n_{\bf p}\leq 1$, there are two classes of  solutions of the variational problem. One class is $\delta n_{\bf p}=0$ with  $n_{\bf p} =0$ or $n_{\bf p} =1$, which corresponds to conventional Fermi liquid. Another class corresponds to the flat band, where $\epsilon_{\bf p}=0$ and  $0< n_{\bf p} <1$. Coexistence of these two classes is illustrated in Fig. \ref{FlatB}.

\begin{figure}
\centerline{\includegraphics[width=0.5\linewidth]{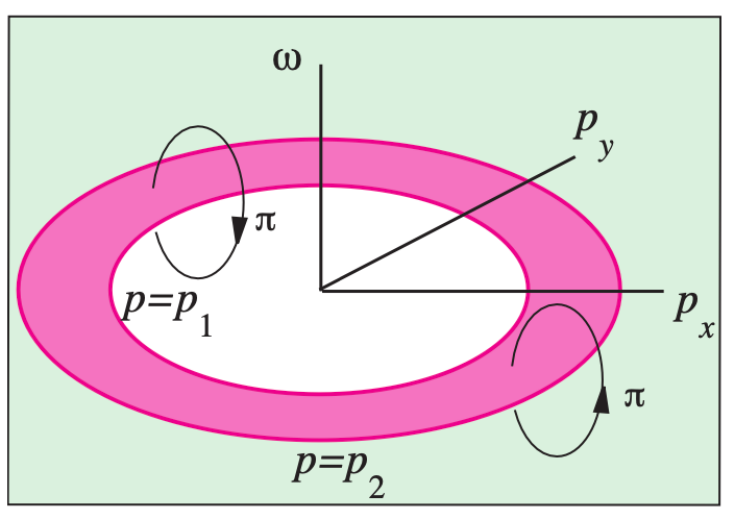}}
\caption{Illustration of the momentum-space topology of the Khodel-Shaginyan flat band on example of the two-dimensional Fermi liquid. The flat band with zero energy occupies the area highlighted in red. The  $2\pi$ winding number of the original Fermi surface splits into the $\pi$ windings about the boundaries of the flat band. This topology is similar to the topology of the Kibble-Lazarides-Shafi cosmic wall bounded by the cosmic strings, or to the topology of splitting of singly quantized vortex to two fractional vortices connected by domain wall.}  
\label{KibbleWall} 
\end{figure}

The Khodel-Shaginyan flat band in Fig. \ref{KibbleWall} emerges as a result of the topological quantum phase transition. In this respect, it represents the momentum-space version of the Kibble-Lazarides-Shafi walls bounded by cosmic strings\cite{Kibble1982,Makinen2019,VolovikComposite2020} -- the composite topological objects described by relative homotopy groups. \cite{Kuang2020} 
There are the other types of the flat bands. Example is the so-called "drumhead" -- the flat band which comes from the topological bulk-surface correspondence. It emerges on the surfaces of semimetals and superconductors with the topologically stable Dirac nodal lines,\cite{Heikkila2011} and on the surface of topological insulators.\cite{NissinenHeikkilaVolovik2021}
 
\subsection{Flat band superconductivity} 
\label{FlatBandSCSec}

Flat bands provide the large density of electronic states, due to which the transition temperature is proportional to the coupling constant instead of the exponential suppression.\cite{KhodelShaginyan1990} For nuclear systems the linear dependence of the gap on the coupling constant has been found by Belyaev.\cite{Belyaev1961} 
That is why the flat band may lead to superconductivity at high temperature, which may even exceed the room temperature. There are several experimental groups, where hints of the room-temperature flat band superconductivity at ambient pressure were detected in graphite systems.\cite{Kawashima2013,Esquinazi2014,Kawashima2018,Nunez-Regueiro2024,Wang2025,Ksenofontov2025}  These experiments suggest that superconductivity takes place at some hidden islands at the surface of the sample, at interfaces or in the columnar objects. The problem is that it is not easy to identify the hidden superconducting island and to determine its structure. Even if a superconducting island is accidentally discovered using a scanning tunneling microscope, it is easily lost, see Ref. \cite{Saunders2018}.

 High $T$ superconductivity at high pressures has been reported in hydrides. Here the important role is played by van Hove singularity, which is close to the Fermi level, see e.g. Refs.\cite{Pickett2016,Polanski2026}. On the other hand, it was shown\cite{Volovik1994,Katsnelson2014} that if the van Hove singularity is sufficiently close to the Fermi level, the interaction between electrons leads to the formation of the flat band by the Khodel-Shaginyan mechanism. This is also in favour of the high-$T$ flat band superconductivity.

\section{Momentum-space topology of topological insulators} 
\label{TopInsulatorsSec}

\subsection{Elasticity tetrads and translation gauge field} 
\label{TranslationSec}

Crystal structure can be considered as a system of three crystallographic surfaces, Bragg planes, of constant phase $X^a(x)=2\pi n^a$, $n^a \in \mathbb{Z}$ with $a=1,2,3$. The intersection of the surfaces
\begin{equation}
X^1({\bf r},t)=2\pi n^1, \,\,  X^2({\bf r},t)=2\pi n^2, \,\, X^3({\bf r},t)=2\pi n^3,
\label{points}
\end{equation}
represent the lattice points of a deformed crystal.
The elasticity theory is described by the elasticity tetrads\cite{DzyaloshinskiiVolovick1980} -- the gradients of the phase functions:
\begin{equation}
 E^{~a}_i(x)= \partial_i X^a(x)\quad i=x,y,z, \quad a=1,2,3, 
\label{reciprocal}
\end{equation}

In the crystalline insulators $N_\omega({\bf p})$ does not depend on ${\bf p}$, being the same for all ${\bf p}$ in the Brillouin zone, $N_\omega({\bf p})\equiv N_\omega=m$, where $m$ is the number of the occupied Brillouin zones.\cite{NissinenHeikkilaVolovik2021} Then there is the connection between this invariant and the invariant 
$\nu_\omega$ in Eq. (\ref{Nofmu}). If the chemical potential $\mu$ is within the band gap,
then $\nu_\omega$ is the total number of fermions in the fully occupied bands of insulator. This can be written in terms of the elasticity tetrads:
 \begin{equation}
\nu_\omega=\frac{1}{(2\pi)^3} N_\omega \int d^3 x \epsilon^{ijk} E^{1}_{i} E^{2}_{j} E^3_{k}
\,,
\label{insulators}
\end{equation}
where the  product of tetrads determines the volume of the Brillouin zone.  

The elasticity tetrads give rise to the topological action in insulators:\cite{NissinenVolovik2019,NissinenHeikkilaVolovik2021}
 \begin{equation}
S_{0D}=\frac{1}{(2\pi)^3} \int d^4x N_\omega\epsilon^{\mu\nu\lambda\rho} E^{1}_{\mu} E^{2}_{\nu} E^3_{\lambda} A_{\rho} \,.
\label{eq:volume_charge}
\end{equation}
Elasticity tetrads enter the topological action in the same way as the $U(1)$ gauge field $A_\mu$, which demonstrate that they play the role of the translational gauge fields.

\subsection{Crystalline insulators and Luttinger theorem} 
\label{InsulatorsSec}

From the topological action (\ref{eq:volume_charge}) one obtains the charge density (or electron density): 
\begin{equation}
J^0=\frac{\delta S}{\delta A_0} =    \frac{1}{(2\pi)^3} N_\omega  \epsilon^{ijk} E^{1}_{i} E^{2}_{j} E^3_{k} \,.
\label{particleDensityInsulator}
\end{equation} 
This corresponds to the volume of the occupied bands, and represents the analog of the Luttinger theorem for insulators: the number density of particles is equal to the volume of the occupied Brillouin zones (here we ignored the spin degeneracy, assuming for simplicity spinless fermions). 
This is the consequence of the general relation between the topological invariant and the number of fermions. The conservation of the topological charge for fermions is equivalent to the conservation of the fermion number. 

This is also true for an interacting electrons. While the real distribution of fermions $n({\bf p})=<a_{\bf p}^\dagger a_{\bf p}>$ is nonzero in the empty bands, the system remains insulating. 

\subsection{Topological invariants in insulators} 
\label{TopInsulatorsSec}

Topological insulators may have different types of the topological terms in the action, such as Chern-Simons, Wess-Zumino terms and $\Theta$ terms. Such terms contain the momentum space topological invariants, gauge fields and elasticity tetrads, see  e.g.  Refs. \cite{NissinenHeikkilaVolovik2021,Hughes2025}.  In the three-dimensional crystals, in addition to the topological term in Eq.(\ref{eq:volume_charge}), there are the actions with different numbers of the elasticity tetrads (dimensionless  prefactors are omitted):
 \begin{eqnarray}
 N_\omega  \int e_{abc} E^a \wedge E^b \wedge E^c \wedge A \,,
\label{S2}
\\
 N^c  \int e_{abc}E^a \wedge E^b  \wedge F \,,
\label{S4}
\\
  N_{a} \int E^a \wedge F \wedge A \,,
\label{S3}
\\
 N_\theta \int F \wedge F  \,.
\label{S1}
\end{eqnarray}

There is a similar chain of topological actions, which contain torsion fields:
\begin{eqnarray}
 \tilde N  \int e_{abc}  E^a  \wedge E^b \wedge T^c \,,
\label{S6}
\\
  N^{c4}  \int e_{abc}T^a \wedge T^b \,.
\label{S5}
\end{eqnarray}
We can see that in addition to the topological invariant $N_\omega$, the crystalline insulators have a large set of topological invariants related to momentum space: $N^c$, $N_a$, $N_\theta$, $\tilde N$ and $N^{c4}$.

For 2-dimensional topological insulators there is a similar set:
\begin{eqnarray}
 N_\omega \int e_{ab} E^a\wedge E^b \wedge A \,,
\label{CS2}
\\
  N_{a} \int E^a\wedge F \,,
\label{CS3}
\\
N \int F \wedge A  \,,
\label{CS1}
\\
N^3  \int e_{ab} T^a\wedge E^b \,.
\label{CS4}
\end{eqnarray}
Here Eq. (\ref{CS2}) describes the topological charge density;
 Eq. (\ref{CS3}) is responsible for the chiral torsional effect of dislocation;\cite{Fujimoto2016,KhaidukovZubkov2018} where the torsion plays the role of effective magnetic field; 
 Eq. (\ref{CS1}) is the Chern-Simons term describing the intrinsic quantum Hall effect; and
Eq. (\ref{CS4}) is related to the Hall effect experienced by dislocation. 
 
 \subsection{On strong CP problem} 
\label{CPSec}

Equation (\ref{S1}) represents the $\Theta$-term in topological insulators. With the dimensionless factors included, this action and the corresponding topological invariant are:\cite{NissinenHeikkilaVolovik2021}   
\begin{align}
S=\frac{N_\theta}{32\pi^2}\int d^4 x \epsilon^{\mu\nu\lambda\rho} F_{\mu\nu} F_{\lambda \rho} \,,\label{eq:theta_term}
\end{align}
\begin{align}
N_\theta &= \frac{1}{96\pi^2}\int_0^{2\pi} du \int d\omega \int_{\rm BZ} d^3 \mathbf{p} \,\epsilon^{u\mu\nu\lambda\rho} \textrm{Tr} \big[(G\partial_u G^{-1}) 
 (G\partial_{\mu}G^{-1})(G\partial_{\nu}G^{-1})((G\partial_{\lambda}G^{-1})(G\partial_{\rho}G^{-1})\big] 
 \,.
 \label{ThetaInvariant}
\end{align}
Here the frequency-momentum space is extended by the periodic adiabatic parameter $u$, and the invariant looks as the Wess-Zumino term in the extended momentum space.
Topological quantization of this action in terms of the momentum-space topological invariant $N_\theta$ suggests the possible solution of the strong $CP$ problem in QCD.
Even if the $CP$ invariance is violated, the topological invariant can remain zero. The value of a topological invariant changes only after a topological quantum phase transition, when the invariant undergoes a jump.

 \section{Conclusion} 
  
  The Landau theory of Fermi liquids is supported by two topological invariants, $N_\omega({\bf p})$ and $N_1$. The first one plays the role of quasiparticle distribution functions, and the second one ensures the topological stability of the Fermi surface.  These invariants confirm the main conjecture of the Landau theory, that the number of  quasiparticles coincides with the number of particles. They  also support the Luttinger theorem, which states, that the volume enclosed by Fermi surface is connected to the particle density. Due to topological stability, the Luttinger theorem remains valid, even if due to electron-electron interaction the pole in Green's function transforms to zero.
  
We also considered the formation of the flat band within the Landau theory of Fermi liquid in the mechanism suggested by Khodel and Shaginyan, and discussed the topology of this flat band.
This mechanism, as well as other topological mechanisms of flat band formation, open the way to room-temperature superconductivity.

The topological consideration is extended to topological insulators, where ,in addition to $N_\omega({\bf p})$, there exists a large set of the momentum-space topological invariants. These induce the topological actions of the Chern-Simons and Wess-Zumino types, which contain the gauge field $A_{\mu}$ and the translational gauge fields in terms of the elasticity tetrads $E^a_\mu$.

\end{document}